\documentclass[12pt]{iopart}
\usepackage{amsfonts,amssymb}
\usepackage{epsfig}
\usepackage{graphicx}
\usepackage{calc}

\usepackage{color}

\begin{document}

\title{Forest-Fire Model with Resistant Trees}
\author{G. Camelo-Neto$^1$ and S. Coutinho$^2$}
\address{$^1$ Laborat\'orio de F\'{\i}sica Te\'orica e Computacional, Universidade Federal de Alagoas, Campus Arapiraca, CEP 57309-005, Arapiraca, Alagoas, Brazil.}
\address{$^2$Laborat\'orio de F\'{\i}sica Te\'orica e Computacional, Departamento de F\'{\i}sica,
Universidade Federal de Pernambuco, CEP 50670-901, Recife, Pernambuco, Brazil.}
\eads{ \mailto{gustavocamelo@arapiraca.ufal.br}, \mailto{sergio@ufpe.br}}

\begin{abstract}
The role of forest heterogeneity in the long-term, large-scale dynamics of forest fires is investigated by means of a cellular automata model and mean field approximation. Heterogeneity was conceived as trees (or acres of forest) with distinct strengths of resistance to burn. The scaling analysis of fire-size and fire-lifetime frequency distributions in the non-interacting fire steady-state limit indicates the breakdown of the power-law behavior whenever  the resistance strength parameter $R$ exceeds a certain value. For \emph{higher} resistant strength, exponential behavior characterizes the frequency distributions, while power-law like behavior was observed for the \emph{lower} resistant case in the same manner as reported in the literature for a homogeneous counterpart model. For the \emph{intermediate} resistance strength, however, it may be described either by a stretched exponential or by a power-law plot whenever the fraction of recovering empty cells by susceptible trees not-exceeds or exceeds a certain threshold respectively, also suggesting a dynamical percolation transition with respect to the stationary forest density.
\end{abstract}
\maketitle

\section{Introduction}
\label{sec:introduction}
A forest-fire model was conceived by Bak, Chen and Tang (BCT)\cite{bak90} in an attempt to study turbulence using the concepts of {\em self-organized criticality} (SOC)\cite{bak87,jensen98}. In such a model, trees are supposed to grow on empty cells at a constant rate $p$ while fire, once started by a single lightning strike, spreads through the system by deterministic local rules. The BCT model however, does not exhibit SOC, but rather a deterministic behavior marked by nontrivial spiral like fire fronts with a well defined length scale ($1/p$) within the limit $p\rightarrow 0$ \cite{grassberger91}. Drossel and Schwabl (DS) \cite{drossel92,clar96} added a new ingredient to the BCT model with the aim of providing a mechanism to ignite small forest clusters in the same manner as the large ones. This new ingredient, the probability of a lightning strike $f$, led to power-law like behavior for both the fire-size (the total number of trees burned in a fire) and the fire-lifetime (the duration of fires in unities of time steps)  frequency distributions, which is considered in the literature the signature of the SOC state within the limit $f/p\rightarrow 0$. However, such power-law distributions have been criticized in the literature\cite{pruessner02,grassberger02}. Nevertheless, the DS forest-fire model has become a paradigm for non-conservative systems exhibiting self-organized criticality\cite{jensen98}.

By the end of the past century some authors began to perform statistics of burned areas of long series of events recorded in historical wildfire database with aim to find signatures of self-organized criticality (power-law) as firstly emphasized in references \cite{malamud98,ricotta99} and later investigated in \cite{song01,turcotte04,malamud05}. However, other frequency distributions of real fires were found in disagreement with a single power-law behavior being smooth but more approximating described by multiple power-law ranges delimited by cutoffs \cite{ricotta01,song01,reed02,telesca05}. Recently, Corral and co-authors \cite{corral08} found that the probability densities of forest-fire waiting times, defined as the time between successive fires above a certain size, verified a scaling law despite the fact that the associated fire-size frequency distribution does not follow a power-law, rather being fitted by a lognormal distribution.

Concerning modeling, more recently the DS model has been explored and generalized by adding structural features and processes to the model in attempt to explain the fire-size multiple power-law or hump-shaped distributions observed in many forest-fire data \cite{lin09}. Other distinct model was also proposed with the same aim by including the main characteristics vegetational growth and fire ignition and propagation \cite{zinck10}.

In the present work, the DS forest-fire model was generalized by introducing heterogeneity into the forested environment, which consists of trees (or acres of forest) with more \emph{resistance} to burn in comparison to the plain \emph{susceptible} tree population. This heterogeneity within the tree populations -- with regard to the strength of resistance to burn -- should be more appropriated for describing fires in natural forests. Moreover, the present simulations also illustrate that such heterogeneity \emph{breaks} the \textit{scaling} behavior that appears in the \textit{homogeneous} DS forest-fire model when the resistance strength to burn reaches a certain threshold. More specifically, the cellular automata (CA) DS forest-fire model \cite{drossel92} was generalized considering the simplest case of a forest composed of two populations of trees -- the \emph{resistant} and the \emph{susceptible} --, and which are randomly distributed on  a two-dimensional square lattice. In addition, each CA-cell is considered to have a neighborhood comprising its eight surrounding cells so called \emph{Moore neighborhood} in contrast with that composed by the first four nearest-neighbors cells (called \emph{von Neumann neighborhood}) adopted by the DS model. Resistant trees are assumed to burn by contact whenever there are at least $R$ burning trees in their neighborhood, while susceptible trees require at least a single neighboring burning tree in order to ignite in the next time step. Hence, $R$ is in some sense the parameter that controls the strength of resistant to burn of the resistant-tree population. Such a deterministic mechanism differs from the stochastic \emph{immunity} previously considered by Drossel and Schwabl~\cite{drossel93}, where the immunity to burn was defined by a certain probability controlling the ignition of a tree surrounded by burning trees. In the present model, the strength of resistance to burn has a deterministic character reinforcing the inherent local inhomogeneities of the distribution of trees along the forested environment.

The coming sections are organized as follows: section \ref{section2} describes de cellular automata model and the limit conditions of the simulations, section \ref{section3} accounts for the study of the fire-size and time-size probability distributions focusing on their long tail behavior more precisely on the breaking of SOC, while section \ref{section4} is dedicated to study of the long term steady-state of the populations with special attention to the dynamical percolation transition. Section \ref{conclusions} summarizes the results and conclusions.

\section{Cellular automata model and simulations}\label{section2}
\subsection{Cellular automata rules and parameters}
The present cellular automata model is defined on a $L \times L$ square lattice with periodic boundary conditions and \emph{Moore} neighborhood.  The variable associated to cellular automaton cells may assume one of the four possible states: \emph{empty} cell ($\mathcal{E}$), \emph{susceptible} tree ($\mathcal{S}$), \emph{resistant} tree ($\mathcal{R}$), or \emph{burning} tree ($\mathcal{B}$). Simulations start from a random configuration composed by susceptible and resistant trees, and empty cells. Afterwards, all cell states are synchronously up-dated according to the following rules:
\begin{enumerate} \itemsep=-1mm
\item[1.] A tree grows on an empty cell with probability $p$, being susceptible with probability $q$ or resistant with probability $(1-q)$.
\item[2.] A susceptible or resistant tree is hit by a lightning strike with probability $f$, becoming a burning tree.
\item[3.] A burning tree dies, becoming an empty cell.
\item[4.] A susceptible tree with at least one burning neighboring tree becomes a burning tree.
\item[5.] A resistant tree with \emph{at least} $R$ burning trees on its neighborhood becomes a burning tree.
\end{enumerate}
One time step is defined as the total time to update in parallel the entire lattice according to rules (1)-(5). Rules (1) and (2) play the role of the environment, which may be associated with factors external to the intrinsic dynamic of fires such as lightning and capacity for growth of trees, while rules (3) through (5) determine the microscopic dynamics of fire propagation. 

The parameter $p$ is the probability that one tree growths on a given empty cell in one time step, while $f$ is the probability that a lightning strikes a single tree ($\mathcal{S}$ or $\mathcal{R}$) in one time step. The latter represents external perturbations, which starts fires (\emph{avalanches}) allowing that small and isolated forest clusters ignite sustaining fire at long-term scale. Both stochastic parameters $p$ and $f$ represent probabilities per time and grid cell and play the same role as the corresponding ones defined in the homogeneous DS model \cite{drossel92,clar96}.

Concerning the tree population heterogeneity the new parameters introduced in the present model was the strength of resistance to burn $R$ and the replenishment parameter $q$. According to the \emph{Moore neighborhood} $R$ may vary from $1$ to $8$. The trivial case $R=1$ characterizes the susceptible-tree $\mathcal{S}$ population (as in the DS model) and $R$ chosen within the interval $2$ to $8$ characterizes the resistant-tree $\mathcal{R}$ population. Therefore the present CA model has two populations of trees: one composed of susceptible trees and the other composed with resistant trees all of them with the same strength of resistant to burn $R$. The parameter $q$ fixed the fraction of new trees that grow on empty cells in the susceptible $\mathcal{S}$ state, hence the $(1-q)$ is the fraction of new trees that grow in  the resistant $\mathcal{R}$ state. The long-term behavior of the forest population when $R$ and $q$ are varied is the focus of interest of this work.

\subsection{Time scales and power-law regime conditions}
The two dynamical characteristic time scales arise through the inverse of probabilities $p$ and $f$. $1/f$ corresponds to the average time between two consecutive lightning strikes, while $1/p$ corresponds to the average time between the birth of two consecutive trees. These two time scales are quite distinct in the limit $f/p\rightarrow 0$, with $f\neq0$. At such a limit the forest has enough time to relax, since a huge amount of trees of order of $p/f$ may grow during the time interval of $1/f$ time steps before a new lightning strike. The limit $p\rightarrow 0$  also  means that fires are extinguished before new trees are able to grow, thus preventing new trees from growing along the edges of a burning cluster before the entire cluster is burned down. On the other hand, the limit $f\rightarrow 0$ means that fires caused by different lightning strikes do not overlap. When considering such a limit regime  the rules of the model can be adapted to reduce the simulation time enormously. Starting from an arbitrary initial configuration composed of empty, susceptible and resistant cells the successive configurations are generated by choosing a cell at random and
\begin{enumerate}
   \itemsep=-1mm
\item[$1'$.] If the cell has a tree ($\mathcal{S}$ or $\mathcal{R}$) the cell becomes a burning tree ($\mathcal{B}$). Fire then spreads over the entire forest cluster to which it belongs, according to the previous rules (3)-(5), being parallel updated until the fire is extinguished. \label{rule1}
\item[$2'$.] If the cell is empty ($\mathcal{E}$),  $p/f$ new cells are chosen at random, assigning to the new empty cell one susceptible tree with probability $q$ or one resistant tree with probability ($1-q$).\label{rule2}
\end{enumerate}
Notice that according to rule $1'$ the fire spreading occurs only within the cluster of trees hit by the lightning, hence the synchronous up-date of the whole lattice becomes restricted to the cluster, saving computation time. Moreover, rule $2'$ introduces a short cut in the time scale contracting the process of replenishment, also saving computation time. Such an alternative particular model was conceived to study the directed percolation model as considered in references \cite{henley93} and \cite{honecker97}. In such a model version, apart from $q$ and $R$, the relevant parameter is $p/f$,  the ratio of the growth and sparking rate \cite{grassberger93}, the limit described above is reached when $p/f\to\infty$.

\subsection{Studied cases}
The above described alternative model was adopted in the present paper to investigate the SOC regime $f/p\rightarrow 0$ with $f\neq0$. The simulations revealed that the system reaches a robust average steady-state configuration after a long transient regardless of the initial configuration. Distinct initial configurations were considered, including the extremes included in all cells at susceptible, resistant, or empty states. The final robust steady state was only reached after different transient times.

Firstly the condition of equiprobable replenishment of trees by susceptible and resistant ones (parameter $q=0.5$) was considered to study the particular cases were $R=$1, 2, 3 and 4. Under such conditions the fire-size and fire-lifetime frequency distributions were calculated after the system reached the steady state, which is characterized by constant average fractions of $\mathcal{E}$, $\mathcal{S}$ and $\mathcal{R}$ cell states.

The emerging results of the simulations reproduced the scaling power-law behavior presented by the fire-size frequency distribution found in the homogeneous DS model whenever the resistance strength $R$=1 and 2. Such behavior found for $R=1$ was expected since in this case both populations are susceptible actually corresponding to a homogeneous forest. However, the power-law behavior found for $R=2$ may be interpreted as resulting from fire propagation in a heterogeneous forested environment but composed by susceptible and \emph{weak} resistant trees. On the other hand, when the resistance strength was chosen $R=$4 and beyond, the overall behavior departed from such a scenario and the cumulative frequency distributions exhibits clear exponential decay, breaking, as expected, the presumed SOC like behavior. However, for the particular case when the strength of resistant of the $\mathcal{R}$ cell population was set equal $R=3$ the system presented a not expected intermediate behavior, the cumulative frequency distribution functions being fitted neither by a power-law or exponential curves. For instance, a stretched exponential function seems to be a good fitting, as presented and discussed in section \ref{section3}.

Secondly, the variation of the replenishment parameter $q$ within the interval $[0,1]$ was considered in the study of the particular cases of tree resistant population with resistant strength $R=$2, 3 or 4. Such study is detailed presented and discussed in section \ref{section4}. It is worth to anticipate that for the intermediate case $R=3$ the fire-size frequency distribution tails can exhibit power-law decay whenever the fraction $q$ of replenishment by $\mathcal{S}$ trees exceed a certain threshold $q_c\simeq 0.82$. Such a change in the cumulative fire-size frequency distribution is followed by a change in the absorbing density steady state indicating the occurrence of a  dynamical percolation phase-transition.

\section{Fire-size and fire-lifetime frequency distributions} \label{section3}

\begin{figure}
\begin{center}
\includegraphics*[width=0.9\columnwidth]{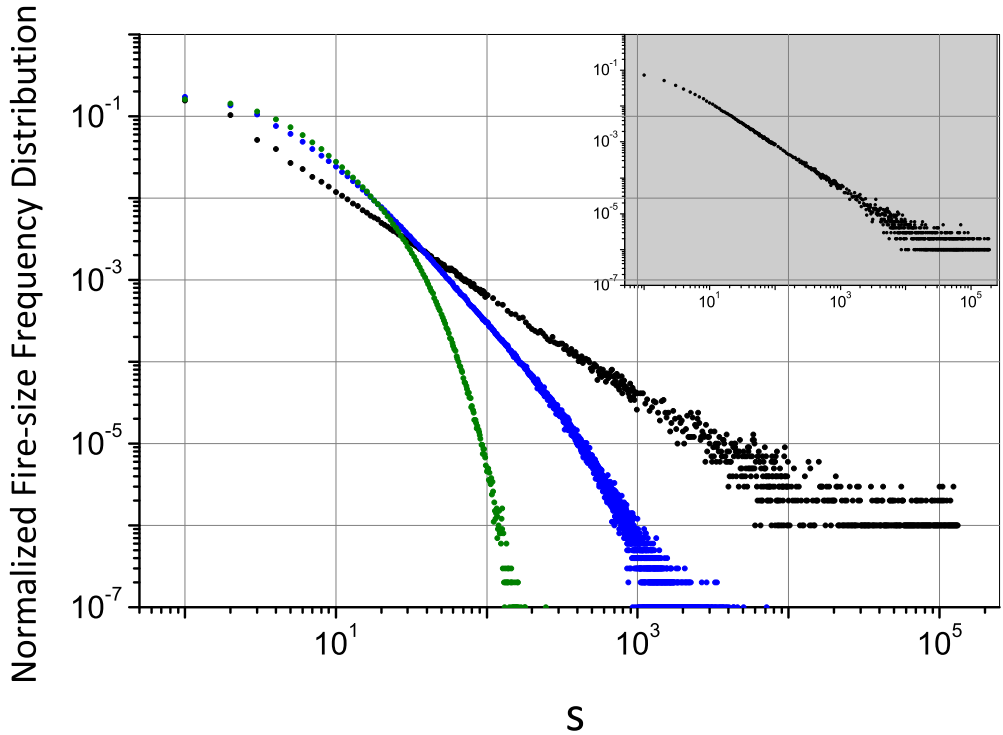}
\caption{(color on-line) Log-Log plot of the normalized fire-size frequency distributions for: weak resistant case $R=2$ (black), intermediate resistant case $R=3$ (blue) and strong resistant case $R=4$ (green). Inset shows the homogeneous $R=1$ case. Simulation parameters: $p/f=10^5$, $q=0.5$. $L=10^4$, $10^6$ samples for $R=1$ and $2$ cases, and $L=2\times 10^4$, $10^7$ samples for $R=3$ and $4$ cases. Many points were skipped in the $R=1$ and $2$ plots to avoid heavy figure file.}
\label{figure01}
\end{center}
\end{figure}

Simulations were carried out using the four-tap algorithm described by Ziff~\cite{ziff98} and suggested by Grassberger~\cite{grassberger93} as the random number generator. The lattice sizes and number of fires however, were restricted by the CPU-time limitations, being $L=10,000$ for $R=1$ and $R=2$, and $L=20,000$ for $R \geq 3$ cases, both with $p/f=10^5$.  Periodic boundary conditions (\emph{torus} conditions) were adopted.  To save computer time, the initial configurations were chosen with 50\% susceptible trees and 50\% resistant chosen at random. Once the system had reached the steady-state, the size and lifetime of each of the $N$ fires were measured. The frequency distribution for fire-size $p(s)$ ($s$ being the \emph{size} of the fires) and fire-lifetime $p(t)$ ($t$ being the fire duration)  were estimated by the corresponding frequency histograms. Figure \ref{figure01} illustrates the \emph{normalized} frequency distribution of fire sizes for $p/f=10^5$ and $R=1-4$. It can be observed that the plots for $R=1$ and $2$ display a power-law like behavior within almost three decades indicating that such a frequency distributions for the fire-size have long tails.

In order to avoid strong fluctuations of the larger sizes caused by poor statistics, the normalized \emph{cumulative} frequency distributions $P(s)=\int_s^\infty p(s')\,ds'/\int_1^\infty p(s')\,ds'$ and $P(t)=\int_t^\infty p(t')\,dt'/\int_1^\infty p(t')\,dt'$ was also investigated.

\begin{figure}
\begin{center}
\includegraphics*[width=0.9\columnwidth]{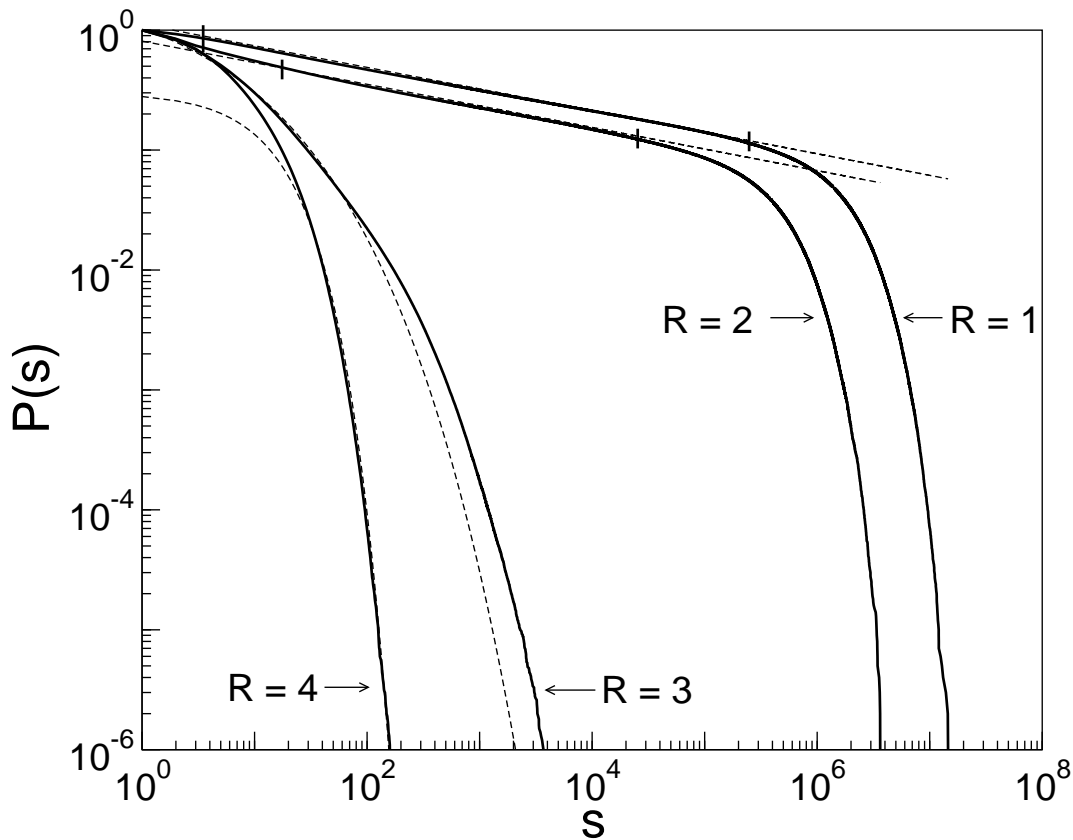}
\caption{Cumulative fire-size distributions for $R=1,2,3$ and $4$.
Simulation parameters are: $L=10^4$ for $R=1$ and $R=2$; $L=2\times 10^4$ for $R=3$ and $R=4$; $p/f=10^5$, and $q=0.5$. Dashed lines indicate the corresponding fittings.}
\label{figure03}
\end{center}
\end{figure}

Figure \ref{figure03} shows the plots of the cumulative fire-size frequency distribution $P(s)$ for forests with a resistance strength of $R=1-4$, considering the fraction of replenishment of susceptible trees as $q=0.5$. Dependence on such a parameter is discussed later on section \ref{section4}. The plot for $P(t)$ have similar characteristics and was skipped in the present manuscript.

In the DS forest-fire model the relevant physical quantities are the fire-size and fire-lifetime frequency distributions. However,  the critical behavior is expected when $p/f\rightarrow\infty$ in the thermodynamic limit $L\rightarrow\infty$. It is necessary, though, to take both limits, $L\rightarrow\infty$, first, and $p/f\rightarrow\infty$ afterwards, hence, the finite-size scaling analysis needs to be done in two variables, $L$ and $p/f$. This feature was well studied in the literature. For instance, the scaling behavior shown by the DS forest-fire model reported in the literature (e.g.  references \cite{pruessner02,grassberger02,satorras00}) is recovered in the present model when $R=1$ is established, in spite of considering the Moore neighborhood. For such a case the system represents a homogeneous forest, since only susceptible trees are considered, the $q$ parameter becoming irrelevant. The plot for $R=2$ displays the same qualitative behavior as that for $R=1$, i.e. slowly decaying cumulative fire-size and fire-lifetime distributions are observed in accordance with (controversial) power-law $P(s) \sim s^{1-\tau}$ and $P(t) \sim t^{1-\beta}$ behavior reported for the DS forest-fire model. Exponents $\tau$ and $\beta$ can be estimated by direct linear fitting. We found $\tau=1.18126(3)$ for $R=1$ and $\tau=1.18432(4)$ for $R=2$ when the scaling intervals are chosen by eyes, as shown in Figure~\ref{figure03}. Analogous results may be obtained for the $\beta$ exponent. Such exponents can be also estimated by the method of moments, introduced by De Menech {\em et al}~\cite{menech98,tebaldi99}, which has the advantage to consider the whole frequency distribution instead of choosing arbitrarily the cutoff by eyes. Such procedure was considered to study the homogeneous DS model in \cite{satorras00}. Adopting this procedure, the following scaling functions and associated exponents are defined for the cumulative fire-size and fire-lifetime frequency distributions, respectively,
\begin{equation}\label{eq01}
P(s)=s^{1-\tau}g\left(\frac{s}{(p/f)^\lambda}\right)\mbox{,}
\end{equation}
and
\begin{equation}
\label{eq02}
P(t) = t^{1-\beta}h\left(\frac{t}{(p/f)^\phi}\right)\mbox{,}
\end{equation}
where $g(x)$ and $h(x)$ are the corresponding scaling functions. 
The rate $p/f$ has to go to infinity as $L^{1/\lambda}$. 

In the present study, the main motivation is to understand how heterogeneity can affect the large scale behavior of fires, so simulations were performed in lattices with fixed large enough sizes to accomplish this goal, doing the finite-size scaling analysis in the variable $p/f$ in the range $10^2$ to $10^5$, similar to the method developed in reference \cite{menech98,tebaldi99}.
Concerning the cases where power-law behavior were observed, Table \ref{table01} displays the estimates of the scaling exponents $\tau$ and $\lambda$, $\beta$ and $\phi$ defined by equation \ref{eq01} for the present model with $R=1$ and $R=2$, as well as the corresponding exponents reported in reference \cite{satorras00} for homogeneous DS model, showing good agreement.
\begin{table}[ht]
\par
\begin{center}
\begin{tabular}{|c|ccc|}
& $R=1$\footnotemark[1] & $R=2$\footnotemark[1] &  DS model\footnotemark[2] \\  \hline
$\tau$    & $1.098$  & $1.069$ & $1.08 \pm 0.01$ \\
$\lambda$ & $1.096$  & $1.062$ & $1.09 \pm 0.01$ \\ \hline
$\beta$   & $1.247$  & $1.228$ & $1.27 \pm 0.01$ \\
$\phi$    & $0.579$  & $0.567$ & $0.59 \pm 0.01$ \\
\end{tabular}
\footnotetext[1]{Present work}
\footnotetext[2]{Reference \cite{satorras00}}
\caption{Scaling exponents associated with the fire-size and fire lifetime frequency distribution in the steady state for homogeneous forests ($R=1$) and inhomogeneous forests with low-resistant trees ($R=2$). Simulation parameters were $q=0.5$,  $L=10,000$, and $p/f$ ranging from $10^4$ to $10^5$.}
\label{table01}
\end{center}
\end{table}

The differences between both estimates may arise due to the fact that the scaling method considers the entire distribution, avoiding the choice of arbitrary cut-offs. We remark that independently of the considered method, both $R=1$ and $R=2$ cases display similar power-law behavior to that observed in the homogeneous DS model (with \emph{von Neumann} neighborhood).

When the resistance parameter has become fixed at $R \geq 4$ ($q=0.5$), it is unlikely that resistant trees will ignite by contact. The exponential distribution of frequencies emerges clearly when the graph is performed on mono-log scale. We also checked visually through snap-shootings of the steady state that the forest evolves to a configuration with most resistant trees forming compact groves, which prevent the occurrence of large fires. In practice, they become burning trees only when directly hit by a lightning strike. As the system evolves, the entire population of trees becomes resistant to fire and the fire-size decreases. An exponential decay in the cumulative fire-size frequency distribution following
\begin{equation}\label{eq03}
P(s) \sim \exp{(-s/s_0)}, \ \ s_0 = 12.36
\end{equation}
was found. The cumulative fire-size frequency distributions $P(s)$ for forests with resistance strength $R>4$ are almost all the same as the case $R=4$ and were omitted for the present discussion.

The intermediate case occurs when $R=3$ ($q=0.5$) where resistant trees are not so strong, but can still defeat fire. In such a case, neither power-law nor exponential behavior were observed.  To understand the behavior of the case $R =3$ and how it differs from the case $R=1,\,2$ (power law) or the $R\geq 4$ cases (exponential tails) we observed that in the former case groves of susceptible trees tends to a sparse fractal structure as $q \rightarrow q_c$, as we noted in from snap-shootings of the steady state. When lightning hits a fractal cluster of susceptible trees, fire should spread in an anomalous manner. The fire front within a cluster marks the boundary of the diffusion of many random self-avoiding walker (SAW) starting from the same point on a given given realization. For clusters with fractal structures close to the one of the percolation cluster the fire front (SAW's) explore the labyrinthine geometry of the giant cluster leading to a slow relaxation regime.  Therefore, the long fractal tails may increase the probability of fires reaching sites from different sides at once, increasing the probabilities of larger fires. Stretch exponentials have been recently used to correct describe the relaxation function for fractal time random walk process in one dimension \cite{aydiner07}.  In the present $R=3$ case, the $P(s)$ distribution may be fitted for instance with a stretched exponential function, i.e.
\begin{equation}\label{eq04}
P(s)=A \exp[-(s/s_0)^\kappa]
\end{equation}
with $A=2.75\pm 0.01$, $s_0=1.08$ and $\kappa=0.36$. The behavior of the fire lifetime frequency distributions are qualitatively identical. The $\kappa$ value obtained is in good agreement with the value $1/3$, expected to relaxation processes that occur when the dilution of the support set reaches the percolation concentration of the underlying lattice.
Such general behaviour was observed in the relaxation of the memory function in a random walks on the diluted hypercube, which was accurately parametrized by a stretched exponential over several orders of magnitude\cite{almeida00,almeida01}.

\section{Time evolution and steady-state regime} \label{section4}
\subsection{Long-term behavior of forest density}
This section is devoted to investigate the role of the replenishment parameter $q$ on the long-term behavior of the trees and empty cells population densities. The time evolution patterns of such population densities vary according to the strength of resistance to burn.

\begin{figure}
\begin{center}
\includegraphics*[width=0.9\columnwidth]{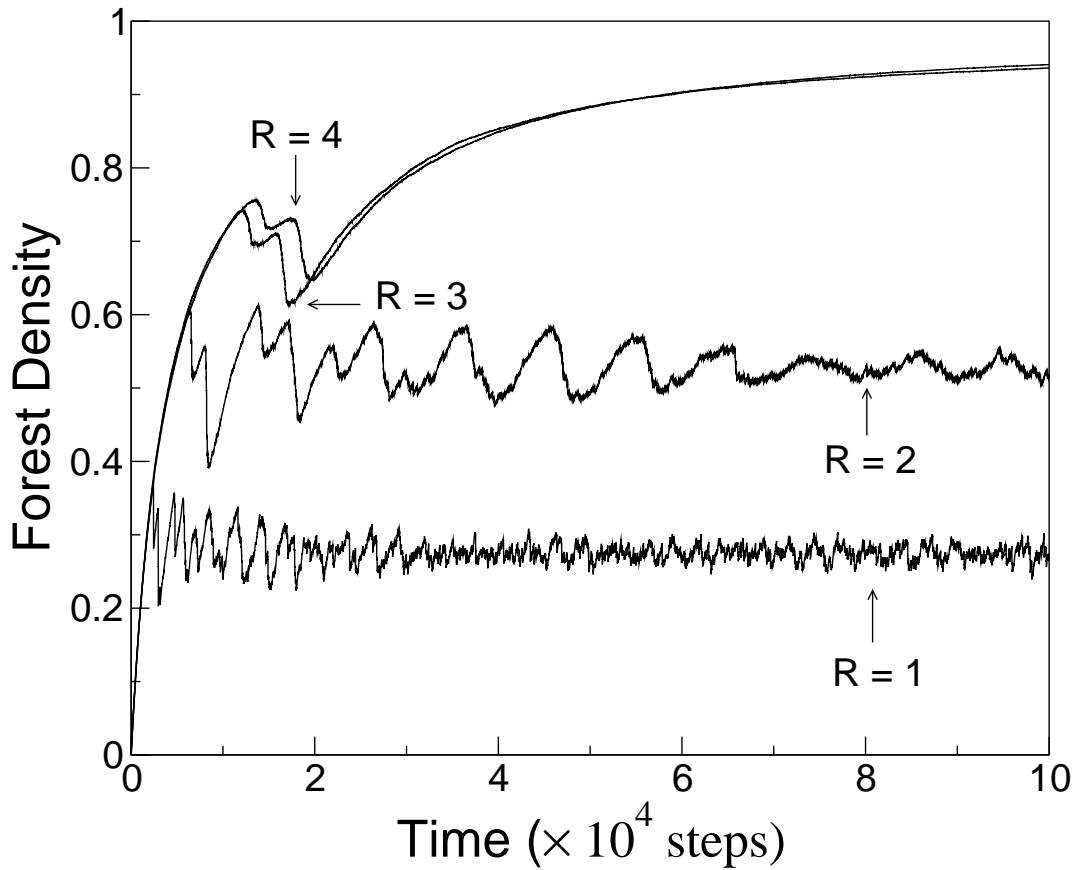}
\caption{Time evolution of the forest density for four different values of $R$. Parameters are: $p/f=10^5$, and $q=0.5$. Lattice size $L=2\times 10^4$.} \label{figure04}
   \end{center}
   \end{figure}

Figure \ref{figure04} displays the time-evolution patterns of the forest density (susceptible + resistant trees) for four different values of $R$ and replenishing parameter $q=0.5$. We note that the time scale is measured in units of $1/f$, and the densities are estimated just after the fires have been extinguished. Therefore, the plots exhibit curves with discontinuities everywhere, the drops representing the fires \emph{avalanches}, whose cumulative fire-size frequency distributions
were displayed in Figure \ref{figure03}.

When analyzed on a coarse-grained scale the \emph{scaling} cases $R=1,\,2$ exhibit highly fluctuating patterns, with stronger fluctuation in the $R=1$ case than the $R=2$. The transient period of resistant cases $R\geq 3$ however, exhibit few strong oscillations at the beginning of the process followed by a smooth increase leading to a high level of the steady state concentration. For the sake of comparison, the initial configuration in all simulations was taken to be the only one with all empty cells. Therefore, the early stage of the dynamics is strongly non-linear, characterized by the continuous growth of the forest up to a critical density followed by a cascade of large fires. After this, forest replenishment resumes, and the process repeats itself  until it achieves the corresponding dynamic equilibrium. However, in all plots, after the characteristic transient period, the forest density would evolve towards the corresponding steady state mean value. The \emph{scaling} cases exhibit strong dispersion (not completely seen in the patterns of Figure \ref{figure04}), while resistant cases display a smooth increase. It is worth noticing however, that the average steady state densities are independent of the initial configuration in all studied cases and depend upon $R$ and $q$. Nevertheless, the long transient periods depend upon such parameters, the size of the lattice and the initial configurations.

\subsection{Equilibrium steady-state population densities}

\begin{figure}
\begin{center}
\includegraphics*[width=0.75\columnwidth]{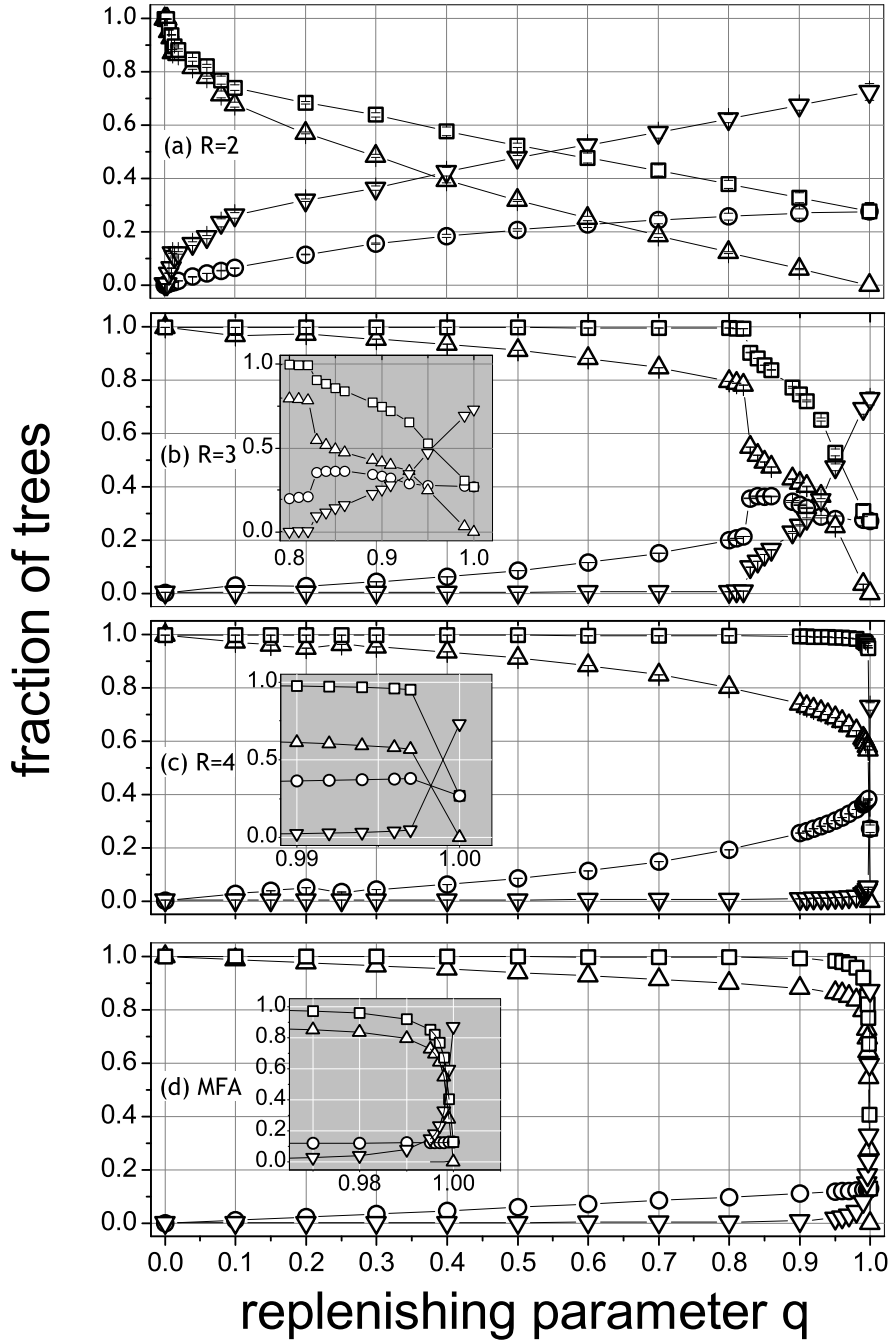}
\caption{Steady-state density of trees and empty cells as a function of the replenishing parameter $q$, for simulations with parameters $p/f=10^5$, $L=10^4$ averaged over $10^3$ events (error bars are smaller than the symbols as illustrated). Panels (a) $R=2$, (b) $R=3$, (c) $R=4$ , and (d) SOC limit in \emph{mean-field approximation}. Insets in panels (b), (c) and (d) illustrate the amplification of the critical regions. Symbol labels: $\triangle$ resistant trees, $\bigcirc$ susceptible, $\bigtriangledown$ empty cells , and $\square$ forest density. Open symbols for $0\leq q<1$ and solid ones for the exact  $q=1$ limit case.  In all cases the steady-state was reached by an automatic procedure based on the mean slope, averaged over $10^3$ simulations.}
\label{figure05}
\end{center}
\end{figure}

The above results shown in Figure \ref{figure04} were obtained considering the replenishment parameter $q=0.5$. The role of such a parameter on the steady-state forest composition was further investigated. The population density of each tree state and empty cells were calculated in the steady-state regime, varying the replenishment parameter $q$ in the interval $[0,1]$ for the resistance strength $R=$2, 3 and 4.
The plots of these population densities as a function of the replenishment $q$ are illustrated in the first three panels of Figure~\ref{figure05} for the heterogeneous forests, corresponding to three distinct scenarios, which can be discriminated according to the resistance strength parameter. 
In all scenarios, however, when $q=0$ the density of trees reaches a trivial absorbing state, in which all cells are occupied by resistant trees, while for $q=1$ the steady state is composed by patches of susceptible trees occupying a fraction $\simeq 0.27\pm 0.02$ of cells, the remanning being empties. Such latter state also corresponds to the steady-state density limit of the \emph{homogeneous} forest model ($R=1$ case) with \emph{Moore} neighborhood. For the corresponding DS model with \emph{von Neumann} neighborhood  the equilibrium forest density reaches $\simeq 0.4084 \pm 0.0001$ \cite{honecker97}.

For $R=2$, an abrupt decay of the forest density is observed as soon as $q$ departs from zero reaching an equilibrium value that diminishes towards the $q=1$ limit one as illustrated in the panel (a) of Figure \ref{figure05}.

For $R=3$ (intermediate case) the forest density remains close to $1$ as $q$ varies from zero to a critical value of $q_c \approx 0.82$. Such a steady state is characterized by a high concentration of resistant trees leading to large patches, which prevents the propagation of fire.  Within this interval, the density of resistant (susceptible) trees decreases (increases) monotonically with $q$, while the density of the empty cells remains quite close to zero. However, when $q\geq 0.82$ an abrupt decay of the forest density occurs concomitantly with an increase of empty cells and susceptible trees. In the interval $(0.82,1]$ the forest density decays towards the homogeneous $q=1$ (or $R=1$) limit value $0.27\pm 0.02$. In this interval the density of resistant trees decreases to zero while the density of susceptible trees reaches a maximum, followed by a decay to the limit of $0.27\pm 0.02$. Since for $q=1$ only susceptible trees survive in the steady-state. 

For $R\geq 4$, the steady-state equilibrium forest density reaches values almost close to one within the whole interval of values $0\leq q < 1$, but abruptly decaying towards the limit of $0.27\pm 0.02$ as $q\rightarrow 1$, as illustrated in Figure \ref{figure05} panel (c).

\subsection{Mean-field approach}

The mean field approximation (MFA) has been applied to investigate the dynamics and steady state of propagation models, such as the forest fire\cite{christensen93,linow06,bancal10} and epidemiological models\cite{bancal10,boguna02,satorras01} in complex networks. In this framework,  the steady-state equilibrium forest density of the cellular-automata model defined by the above rules (1)-(5) is studied following the approach defined in reference \cite{camelo96} for a similar CA model. 

The fractions of each cell state $D_x=D_x$ ($x=\mathcal{S},\,\mathcal{R},\,\mathcal{E}$ or $\mathcal{B}$) can be  obtained from the coupled difference equations for $\Delta D_x = D_x(t+1)-D_x(t)$, i.e.
\begin{eqnarray*}
  \Delta D_\mathcal{E} &=& D_\mathcal{B}(t)-p\, D_\mathcal{E}(t)   \\
  \Delta D_\mathcal{S} &=& q\, p\, D_\mathcal{E}(t)-[\Theta+f(1-\Theta) ]\,D_\mathcal{S}(t)  \\
  \Delta D_\mathcal{R} &=& (1-q)\,p\,D_\mathcal{E}(t)-[\Phi_R+f(1-\Phi_R)]\, D_\mathcal{R}(t) \\
  \Delta D_\mathcal{B} &=& [\Theta\, D_\mathcal{S}(t) + \Phi_R\, D_\mathcal{R}(t)]+ f\,[(1-\Theta)\,D_\mathcal{S}(t)+\nonumber \\
  && + (1-\Phi_R)\,D_\mathcal{R}(t)] -D_\mathcal{B}(t)
\end{eqnarray*}
where $\Theta(t)= 1-(1-D_\mathcal{B}(t))^8$ is the probability that a $\mathcal{S}$ cell has at least one surrounding $\mathcal{B}$ cell in the time step $t$ and
\begin{equation*}
\Phi_R(t)=\sum_{n=R}^8 \left(_n^8\right)\,D_\mathcal{B}(t)^n(1-D_\mathcal{B}(t))^{8-n}
\end{equation*}
is the probability that a $\mathcal{R}$ cell has at least $R$ neighboring $\mathcal{B}$ cells at time step $t$. Such densities of cells must obey the normalization condition at each time step,
\begin{equation*}
    D_\mathcal{S}(t)+D_\mathcal{R}(t)+D_\mathcal{B}(t)+D_\mathcal{E}(t)=1
\end{equation*}
In the steady-state $D_x \rightarrow \tilde{D}_x$, that is, $\Delta \tilde{D}_\mathcal{E}=\Delta \tilde{D}_\mathcal{S}=\Delta \tilde{D}_\mathcal{R}=\Delta \tilde{D}_\mathcal{B}=0$, hence
\begin{eqnarray*}
  \tilde{D}_\mathcal{E} &=& \frac{1}{p}\,\tilde{D}_\mathcal{B}  \\
  \tilde{D}_\mathcal{S} &=& \frac{q\,\tilde{D}_\mathcal{B}}{(1-f)\Theta+f} \\
  \tilde{D}_\mathcal{R} &=& \frac{(1-q)\,\tilde{D}_\mathcal{B}}{(1-f)\Phi_R+f}
\end{eqnarray*}
where $\tilde{D}_\mathcal{B}$ is given by the solutions of
\begin{equation*}
    \tilde{D}_\mathcal{B}\,\left[1+\frac{1}{p}+\frac{q}{(1-f)\Theta+f}+\frac{(1-q)}{(1-f)\Phi_R+f}\right]=1
\end{equation*}
with $\Theta$ and $\Phi_R$ being their corresponding values at the steady state.

A similar plot can be drawn for the SOC limit condition studied here, say $p\rightarrow 0$, $f \rightarrow 0$ with $f/p \rightarrow 0$. For $0 \leq q<1$ the fractions of burning trees $\tilde{D}_b$, empty cells $\tilde{D}_\mathcal{E}$, resistant $\tilde{D}_\mathcal{R}$ and susceptible $\tilde{D}_\mathcal{S}$ trees are given by
\begin{eqnarray}
\tilde{D}_\mathcal{B}& \simeq &\frac{X(q)}{(1-q)}\,f \rightarrow0, \\
\tilde{D}_\mathcal{E}&=&\frac{X(q)}{(1-q)}\,\frac{f}{p} \rightarrow0,\\
\tilde{D}_\mathcal{R}&=& X(q),  \\
\tilde{D}_\mathcal{S}&=&\frac{7\,q\, X(q)}{56\,X(q)+16(1-f)}\\
X(q)&=&\frac{7}{16}[1+\sqrt{1+32\,(1-q)/49}]
\end{eqnarray}
For $q=1$ however, the MFA-limit densities are $0$, $7/8$ and $1/8$ for the resistant, empties and susceptible trees respectively, in slight numerical disagreement with the results obtained by simulation. The MFA results are displayed in the panel (d) of figure~\ref{figure05}. It is worth to mention that MFA-SOC limit is independent of the resistance strength, $R$, as expected from the absence of local correlations in the MFA.

\section{Concluding remarks}\label{conclusions}

The present work investigates the role of heterogeneity in the previous homogeneous forest-fire model reported in the literature. Heterogeneity was considered by introducing a population of trees with a certain strength of resistant to burn defined by a local interaction and by a certain rate of replenishment. The main result emerging from the present study suggests that such ingredients play an essential role in the propagation process of forest-fires, changing the long-term, large-scale behavior of the overall phenomenon. The combination of both ingredients may lead the systems to reach two scenarios: a steady state characterized by \emph{power-law} (long tail) fire-size and fire-lifetime frequency distributions, which is identified in the literature with the so-called \emph{self-organized critical} state, or to converge to an \emph{exponential} decay of such distribution, the latter leading to a forested environment composed by a background of resistant trees with patches of weak-resistant susceptible trees. For certain critical values of the parameters governing the heterogeneity the system steady state may undergoes a dynamical phase transition between absorbent states characterizing such scenarios. 

Real forests are very complex non-equilibrium, non-linear systems and many factors contribute to heterogeneities across the landscape of forested land. A forest in equilibrium can be characterized by having a species composition with age structure unchanged for long periods of time. Probably, few temperate forests may be found near an equilibrium state, since fire, recurring pests, human interventions and other factors all tend to drive the system to a stochastic equilibrium. Data from real forest fires have been collected in many countries over the past decades and analyzed under many aspects.  Regarding the frequency distribution of size of fires, many authors observed a clear power law behavior, while others have found deviations from this behavior but without doing a careful analysis of the diversity, age structure and other special characteristics of tree populations in the forests studied. 
The results of this study suggest that deviations in the frequency distribution of fire-sizes observed in many forests could be due to inhomogeneities in the populations of trees with respect to their resistance strength to ignition and its rate of replenishment. Further, such distinct populations evolve to a critical state of stochastic equilibrium. Further studies to characterize the nature of this transition is now being considered.

\ack
We grateful acknowledge the enlightening discussion and comments of Barbara Drossel and her thoughtful critical reading of the manuscript. G. Camelo-Neto particularly acknowledge to Thiago G. dos Santos who helped to obtain some plots of figure \ref{figure05}. This work received financial support from CNPq and CAPES (Brazilian federal grant agencies), from FACEPE (Pernambuco state grant agency under the grants PRONEX EDT 0012-05.03/04 and PRONEX APQ 0203-1.05/08), and from FAPEAL (Alagoas state grant agency under the grant 35.0114/2006-4).


\begin{thebibliography}{10}
\expandafter\ifx\csname url\endcsname\relax
  \def\url#1{{\tt #1}}\fi
\expandafter\ifx\csname urlprefix\endcsname\relax\def\urlprefix{URL }\fi
\providecommand{\eprint}[2][]{\url{#2}}

\bibitem{bak90}
Bak P, Chen K and Tang C 1990 {\em Physics Letters A\/} {\bf 147} 297--300

\bibitem{bak87}
Bak P, Tang C and Wiesenfeld K 1987 {\em Phys. Rev. Lett.\/} {\bf 59} 381--384

\bibitem{jensen98}
Jensen H~J 1998 {\em Self Organized Criticality - Emergent Complex Behavior in
  Physical and Biological Systems\/} (Cambridge University Press, New York)

\bibitem{grassberger91}
Grassberger P and Kantz H 1991 {\em Journal of Statistical Physics\/} {\bf 63}
  685 -- 700

\bibitem{drossel92}
Drossel B and Schwabl F 1992 {\em Phys. Rev. Lett.\/} {\bf 69} 1629--1632

\bibitem{clar96}
Clar S, Drossel B and Schwabl F 1996 {\em Journal of Physics: Condensed
  Matter\/} {\bf 8} 6803--6824

\bibitem{pruessner02}
Pruessner G and Jensen H~J 2002 {\em Phys. Rev. E\/} {\bf 65} 056707

\bibitem{grassberger02}
Grassberger P 2002 {\em New Journal of Physics\/} {\bf 4} 17
  \urlprefix\url{http://stacks.iop.org/1367-2630/4/17}

\bibitem{malamud98}
Malamud B, Morein G and Turcotte D 1998 {\em Science\/} {\bf 281} 1840--1842

\bibitem{ricotta99}
Ricotta C, Avena G and Marchetti M 1999 {\em Ecological Modelling\/} {\bf 119}
  73--77

\bibitem{song01}
Song W, Weicheng F, Binghong W and Jianjun Z 2001 {\em Ecological Modelling\/}
  {\bf 145} 61--68

\bibitem{turcotte04}
Turcotte D~L and Malamud B~D 2004 {\em Physica A: Statistical Mechanics and its
  Applications,\/} {\bf 340} 580

\bibitem{malamud05}
Malamud B~D, Millington J~D~A and Perry G~L~W 2005 {\em Proceedings of the
  National Academy of Sciences of the United States of America\/} {\bf 102}
  4694--4699

\bibitem{ricotta01}
Ricotta C, Arianoutsou M, Diaz-Delgado R, Duguy B, Lloret F, Maroudi E,
  Mazzoleni S, Moreno J~M, Rambal S, Vallejo R and Vazquez A 2001 {\em
  Ecological Modelling\/} {\bf 141} 307--311

\bibitem{reed02}
Reed W~J and McKelvey K~S 2002 {\em Ecological Modelling\/} {\bf 150} 239--254

\bibitem{telesca05}
Telesca L, Amatulli G, Lasaponara R, Lovallo M and Santulli A 2005 {\em
  Ecological Modelling\/} {\bf 185} 531--544

\bibitem{corral08}
Corral A, Telesca L and Lasaponara R 2008 {\em Physical Review E\/} {\bf 77}
  016101

\bibitem{lin09}
Lin J and Rinaldi S 2009 {\em Ecological Modelling\/} {\bf 220} 898--903

\bibitem{zinck10}
Zinck R~D, Johst K and Grimm V 2010 {\em Ecological Modelling\/} {\bf 221} 98
  -- 105 ISSN 0304-3800 special Issue on Spatial and Temporal Patterns of
  Wildfires: Models, Theory, and Reality
 

\bibitem{drossel93}
Drossel B and Schwabl F 1993 {\em Physica A: Statistical Mechanics and its
  Applications\/} {\bf 199} 183 -- 197 ISSN 0378-4371

\bibitem{henley93}
Henley C~L 1993 {\em Phys. Rev. Lett.\/} {\bf 71} 2741--2744

\bibitem{honecker97}
Honecker A and Peschel I 1997 {\em Physica A: Statistical and Theoretical
  Physics\/} {\bf 239} 509--530

\bibitem{grassberger93}
Grassberger P 1993 {\em Journal of Physics A: Mathematical and General\/} {\bf
  26} 2081--2089

\bibitem{ziff98}
Ziff R~M 1998 {\em Computers in Physics\/} {\bf 12} 385 -- 392

\bibitem{satorras00}
Pastor-Satorras R and Vespignani A 2000 {\em Phys. Rev. E\/} {\bf 61}
  4854--4859

\bibitem{menech98}
De~Menech M, Stella A~L and Tebaldi C 1998 {\em Phys. Rev. E\/} {\bf 58}
  R2677--R2680

\bibitem{tebaldi99}
Tebaldi C, De~Menech M and Stella A~L 1999 {\em Phys. Rev. Lett.\/} {\bf 83}
  3952--3955

\bibitem{aydiner07}
Ekrem Aydiner, 2007, {\em Chin. Phys. Lett.\/}{\bf 24} 1486-1489  

\bibitem{almeida00}
de~Almeida R~M~C, Lemke N and Campbell I~A 2000 {\em European Physical Journal
  B\/} {\bf 18} 513

\bibitem{almeida01}
de~Almeida R~M~C, Lemke N and Campbell I~A 2001 {\em Journal of Magnetism and
  Magnetic Materials\/} {\bf 226} 1296

\bibitem{christensen93}
Christensen K, Flyvbjerg H and Olami Z 1993 {\em Phys. Rev. Lett.\/} {\bf 71}
  2737--2740

\bibitem{linow06}
{M\"uller-Linow, Mark and Marr, Carsten and H\"utt, Marc-Thorsten} 2006 {\em
  Phys. Rev. E\/} {\bf 74} 016112

\bibitem{bancal10}
Bancal J~D and Pastor-Satorras R 2010 {\em The European Physical Journal B -
  Condensed Matter and Complex Systems\/} {\bf 76}(1) 109--121 ISSN 1434-6028
  10.1140/epjb/e2010-00165-7
  \urlprefix\url{http://dx.doi.org/10.1140/epjb/e2010-00165-7}

\bibitem{boguna02}
Bogu\~n\'a M and Pastor-Satorras R 2002 {\em Phys. Rev. E\/} {\bf 66} 047104

\bibitem{satorras01}
Pastor-Satorras R and Vespignani A 2001 {\em Phys. Rev. E\/} {\bf 63} 066117

\bibitem{camelo96}
Camelo-Neto G and Coutinho S 1996 {\em Fractals\/} {\bf 4} 113--122

\end{thebibliography}


\section*{References} 

\providecommand{\newblock}{}

\end{document}